**The role of magnetic and orbital ordering at the metal-insulator transition in NdNiO$_3$.**

V. Scagnoli[1], U. Staub[1], A. M. Mulders[1], M. Janousch[1], G. I. Meijer[2], G. Hammerl[2], J. M. Tonnerre[3], and N. Stojic[4]

[1]Paul Scherrer Institut, 5232 Villigen PSI, Switzerland

[2]IBM Research, Zurich Research Laboratory, 8803 Rüschlikon, Switzerland

[3]Laboratoire de Cristallographie, CNRS, 38042 Grenoble, France

[4]Abdus Salam International Centre for Theoretical Physics, Trieste 34014, Italy

Soft x-ray resonant scattering at the Ni $L_{2,3}$ edges is used to test models of magnetic- and orbital-ordering below the metal-insulator transition in NdNiO$_3$. The large branching ratio of the $L_3$ to $L_2$ intensities of the (1/2 0 1/2) reflection and the observed azimuthal angle and polarization dependence originates form a non-collinear magnetic structure. The absence of an orbital signal and the non-collinear magnetic structure show that the nickelates are materials for which orbital ordering is absent at the metal-insulator transition.

The $RE$NiO$_3$ compounds ($RE$ is a rare earth or Y ion) have attracted considerable interest as they exhibit a temperature driven metal-insulator (MI) transition for most of the $RE$ ions[1]. These nickelates crystallize in the orthorombically distorted perovskite structure *Pbnm* (a=5.38Å, b=5.39Å and c=7.61Å) with unit cell approximately $\sqrt{2}a_p \times \sqrt{2}a_p \times 2a_p$, where $a_p$ is the cubic perovskite cell parameter. They exhibit a structural phase transition when entering the insulating state.[2,3] For the light $RE$ ions, a simultaneous antiferromagnetic ordering occurs ($T_N=T_{MI}$), whereas for the heavier RE ions $T_{MI}>T_N$.[4] The magnetic ordering proposed by neutron powder diffraction[5] consists of a propagation vector **k**=(1/2 0 1/2) and an unusual up-up-down-down stacking of ferromagnetic planes along the simple cubic [111] direction. This magnetic structure with alternating ferro– and antiferromagnetic neighbours cannot be described by simple antiferromagnetic or ferromagnetic exchange couplings between the spins. The nominal valence of Ni is 3+ and $RE$NiO$_3$ is considered to be self-doped. Therefore, these nickelates are viewed as ideal model compounds because no structural disorder due to chemical substitution is required for doping. However, orbital ordering (OO) can remove the degeneracy of a single occupied $e_g$ electron in a cubic field. This process may lead to OO with the same wavevector as the magnetic ordering and could naturally explain the unusual magnetic ordering wavevector. High-resolution powder diffraction,[6] Raman scattering[7] and resonant hard x-ray scattering[3] have found a symmetry breaking at $T_{MI}$, indicating a charge disproportionation, leading to a ground state with monoclinic symmetry. Resonant x-ray scattering experiments at the Ni $K$ edge have shown[8] that the resonant signal of the unrotated light at the (033) reflection is dominated by charge disproportionation and not caused by the asphericity of the Ni ions. This is in contrast to results of resonant x-ray studies on manganites.[9,10]

Different models for the magnetic structure based on neutron powder diffraction[5,6,11] have been proposed and due to the lack of sizable single crystals, the magnetic structure remains ambiguous. A search for reflections of type ($h$/2 $k$ $l$/2) was performed at the Ni $K$ edge and off-resonance[12] in order to establish if the unusual magnetic ordering is caused by orbital ordering. No intensity has been found at such points in reciprocal space, indicating that the magnetic and possible OO signals are too weak to be observed. The orbital signal is expected to be weak at the Ni $K$ edge, which intensity is commonly dominated by the Jahn-Teller distortion. Indeed the latter has not been detected by neutron diffraction.

Resonant soft x-ray scattering is a very powerful tool to study the magnetic and orbital ordering schemes of transition-metal ions. The 3$d$ states are directly probed by the dipole 2$p$ to 3$d$ transitions ($L$ edges), leading to very strong enhancements in the scattered intensity, independent of the Jahn-Teller distortion.[13] Recently, the first successful resonant soft x-ray scattering experiments on magnetic and orbital ordering in bulk materials were presented.[14-17]

In this report, we present a resonant soft x-ray scattering study on the (1/2 0 1/2) reflection of an epitaxial film of NdNiO$_3$. We performed azimuthal-angle ($\psi$) scans (rotation about the Bragg wavevector) as well as polarization analysis of the scattered beam at the Ni $L_3$ edge. The data are well described by magnetic contributions only. No hint of orbital order with this particular **k** vector is found. The magnetic signal is not consistent with an up-up-down-down spin arrangement and indicates a non-collinear magnetic ordering scheme. Clarification of the magnetic structure corroborates the absence of orbital ordering with a (1/2 0 1/2) wavevector.

Epitaxial films of NdNiO$_3$ were grown on [101] oriented NdGaO$_3$ substrates (*Pbnm*) by pulsed laser deposition as in Ref. 6. The thickness of the film is approximately 500 Å. The resistivity of the film was measured with the conventional 4-probe technique and shows a first-order metal-insulator transition at approximately $T$=200 and 180 K upon heating and cooling, respectively. Polarized soft x-ray scattering experiments were performed at the SIM beamline of the Swiss Light Source at the Paul Scherrer Institut using the RESOXS endstation.[18] Measurements were performed at the Ni $L_{2,3}$ edges between 30 and 300 K using a continuous helium-flow cryostat. Azimuthal scans were obtained by sample rotation with an accuracy of approximately 5°. The linear polarization of the incoming beam could be rotated from horizontal ($\pi$) to vertical ($\sigma$). Polarization analysis of the scattered radiation was performed with a W/C multilayer. Horizontal and vertical mounting of this multilayer allowed for $\pi$' and $\sigma$' detection of the scattered radiation, respectively.

The energy dependence of the intensity of the (1/2 0 1/2) diffraction peak, measured without polarization analysis, is shown in Fig. 1 for both incident polarizations $\pi$ and $\sigma$. The recorded intensities have been corrected for absorption of the 500-Å-thick film using collected electron yield data. Strong resonant enhancements are observed at the $L_3$ and $L_2$ edges. The intensity at the $L_3$ edge is more than an order of magnitude larger than the intensity at the $L_2$ edge. Similar branching ratios have been observed for the magnetic intensities at the Mn $L_{2,3}$ edges of manganites,[14,17] indicative of a general trend for transition metal oxides. Both edges contain two features, a narrow intense peak at low energy and a broad, weaker peak at high energy separated by 1.7 ± 0.1 and 1.4 ± 0.1 eV for the $L_3$ and $L_2$ edges, respectively. The temperature dependence of the (1/2 0 1/2) reflection upon cooling and heating is shown in Fig. 2. A clear hysteresis, consistent with resistivity measurements, is observed, indicative of a first-order phase transition.

A distinction between charge, magnetic, and orbital scattering can only be obtained by measuring azimuthal dependences, as recently shown for soft x-rays at the $L_{2,3}$ edges of Mn in $La_{0.5}Sr_{1.5}MnO_4$.[17] Additionally, identification of the polarization of the incoming and outgoing x-ray intensity is often mandatory for a meaningful interpretation of the data. Polarization analysis for this purpose has only recently been employed for the soft x-ray regime.[17] Because the (1/2 0 1/2) type reflections have not been observed at and below the Ni $K$ edge,[12] a spherical and and an aspherical charge contribution to the scattered intensity can be excluded. To establish an orbital contribution, symmetry analysis is very helpful. However, the proposed symmetry for the orbital ordering $Bb2_1m$[19] contradicts the resonant signal of the ($0kl$) and ($h0l$) type reflections at the Ni $K$ edge.[3] The resonant signal of these latter reflections has been interpreted as originating from charge disproportionation resulting in a monoclinic $P2_1/n$ symmetry as previously proposed by high-resolution powder diffraction.[6]

Figure 3 displays the azimuthal angle ($\psi$) dependence taken at the $L_3$ edge (857.4 eV) for incident $\pi$ and $\sigma$ polarizations. In analogy to the manganates, one would expect the $L_2$ edge to be more sensitive to an orbital contribution. However, azimuthal angle scans at 857.4 eV, ($L_3$), and 874.4 eV and 875.6 eV ($L_2$) show the same $\psi$ dependence within the experimental accuracy. This indicates that the scattering originates from a single process of either magnetic or orbital origin. The ratio $\pi/\sigma$ is greater than unity at any $\psi$. A two-fold periodic signal is observed for incident $\pi$ radiation, whereas for $\sigma$ incident radiation the diffraction intensity is only weakly dependent on $\psi$. The energy dependence of the (1/2 0 1/2) reflection, taken at selected $\psi$, displays a very similar form. In general, the average orbital scattering is expected to be strongest with incident $\sigma$ polarization. In contrast, the magnetic signal shows no scattering in the $\sigma$-$\sigma$ channel and therefore the magnetic signal is expected to be strongest with $\pi$ incident radiation.[20] Since the scattering with $\pi$ incident radiation is strongest (Fig. 1) the (1/2 0 1/2) reflection is likely to be of magnetic origin. To give further support to the magnetic origin of the scattering, polarization analysis was performed at the maximum intensity of the $L_3$ edge as shown in Fig. 4 for $\Psi=90°$. A slightly larger signal is observed in the $\pi$-$\pi$' channel compared to the $\sigma$-$\pi$' channel below $T_{MI}$. The $\pi$-$\pi$' intensity is absent for $T>T_{MI}$ (upper panel, Fig. 4). Additionally, there is a clear signal for $\pi$-$\sigma$' whereas no signal is detected for $\sigma$–$\sigma$ (lower panel, Fig 4). These findings are in excellent agreement with the assumption of a solely magnetic origin of the scattering.

To compare the data with the various proposed magnetic models for $RENiO_3$,[5, 6, 11] the structure factor expressed as[21]

$$F = \sum_{K,Q,q} (-1)^Q H^K_{-Q} D^K_{Qq} \sum_{\mathbf{d}} e^{i\mathbf{d}\cdot\mathbf{\tau}} \langle T^K_q \rangle_{\mathbf{d}}, \qquad (1)$$

was calculated, where $\langle T^K_q \rangle_{\mathbf{d}}$ represents the physical origin of the scattering and is a spherical tensor describing the multipoles of rank $K$ of the $3d$ shell of the Ni ion at position $\mathbf{d}$ and $\mathbf{\tau}$ is a reciprocal lattice vector. $H^K_{-Q}$ describes the experimental geometry, including polarization and $D^K_{Qq}$ is a rotation matrix to transform between the local and experimental coordinate system. For the magnetic model displayed in Fig. 2 (inset) with $Pbnm$ symmetry and $K=1$, the magnetic structure factors for the different polarization channels are

$$\begin{aligned}
F_{\sigma-\sigma'} &= 0 \\
F_{\pi-\pi'} &= -4\sin(2\theta)\sqrt{2}\left[(i-1)(\cos(\gamma_0)\sin(\beta_0)\langle T_x \rangle) + (1+i)(\cos(\beta_0)\langle T_z \rangle)\right] \\
F_{\sigma-\pi'} &= 2\sqrt{2}\{2(1+i)\cos(\theta-\alpha_0)\sin(\beta_0)\langle T_z \rangle + \\
&\quad (1-i)\langle T_x \rangle[\cos(\theta-\alpha_0+\gamma_0)(-1+\cos(\beta_0)) + \cos(\theta-\alpha_0-\gamma_0)(1+\cos(\beta_0))]\}
\end{aligned} \qquad (2)$$

where $\theta$ is the Bragg angle, $\cot(\alpha_0) = \cot(\beta)\sin(\psi)$, $\sin(\beta_0) = \cos(\beta)\cos(\psi)$, $\cot(\gamma_0) = \sin(\beta)\cot(\psi)$, and $\beta = \arctan(a/c)$, with $a$ and $c$ the lattice constants. The structure factor for π–σ' is the same as for σ–π' except that $\alpha_0 \rightarrow -\alpha_0$ and $\gamma_0 \rightarrow -\gamma_0$. The operators $\langle T_x \rangle$ and $\langle T_z \rangle$ represent the Cartesian components of spherical tensors and are proportional to components of the magnetic moments of the Ni ions. Calculation of scattered intensity gives best agreement with the observed $\psi$ dependence (see Fig. 3) for $\langle T_x \rangle \approx \langle T_z \rangle$. Note that there is one overall scale factor for both polarizations in the calculation. Our analysis indicates that the (1/2 0 1/2) reflection is not sensitive to magnetic components along the $b$ axis. A collinear spin arrangement as proposed for NdNiO$_3$,[5] and YNiO$_3$[6] cannot reproduce the azimuthal angle dependence (see Fig. 3). For HoNiO$_3$,[11] both a collinear and a non-collinear model were proposed, but neutron diffraction could not differentiate between these. Also, for monoclinic symmetry, collinear spin arrangement is inconsistent with our data. In this case the magnetic contributions from the two Ni sites may have different energy dependences, due to their different charges.

The observed energy dependence of the (1/2 0 1/2) reflection cannot be understood by an integer valence of Ni$^{2+}$ or Ni$^{3+}$. However, our simulations for the monoclinic phase[21] using the charge transfer (configuration interaction) multiplet approach[22] are consistent with the prediction of an estimated charge valence difference of ~0.45.[6] Calculated spectra were fitted to the experimental magnetic scattering and, as an additional constraint, we requested an agreement with the absorption spectrum in the insulating phase of rare earth RENiO$_3$ compounds.[23] Using a single site (assuming all Ni ions have the same valence) of integer valence Ni$^{3+}$, Ni$^{2+}$, as well as a mixed valence (61% 3d$^7$ + 37% 3d$^8$**L** + 2% 3d$^9$**L**$^2$, which describes absorption well[23]), we were not able to obtain a simultaneous agreement with the scattering and absorption spectra. The scattering spectra of Ni$^{3+}$ and the mixed valence spectra were improving by adding more $3d^8\underline{L}$ component. A simulation including $Ni_A$ and $Ni_B$ sites reproduces magnetic spectrum well (shown in the inset of Fig. 1) and is also consistent with the absorption spectrum. The corresponding configuration interaction parameters are $U_{dd}$ -$U_{pd}$=-2 eV, and the transfer integrals: $T(A_1)=T(B_1)=2.0$ eV and $T(B_2)=T(E)=1.3$ eV, while for $Ni_A$, the crystal field is described by $10D_q=1.8$ eV, $D_s = 0.1$ eV, the charge transfer $\Delta=1.5$ eV and for $Ni_B$ $10D_q=1.5$ eV, $\Delta = -1.7$ eV. The resulting ground state of $Ni_A$ is 64% $3d^7$ + 36% $3d^8\underline{L}$, and the spin and local crystal field axis enclose an angle $\alpha$ of 90°. [21] The $Ni_B$ ground state configuration is composed of 32% $3d^7$+68% $3d^8\underline{L}$ and, since the $Ni_B$ is in a almost cubic crystal field, there is no difference in the calculated spectra for $\alpha=0°$ and 90° for $Ni_B$ spin (and, hence, no difference between a collinear and non-collinear spin structure). Nonetheless, we can identify the contribution to the magnetic scattering from $Ni_A$ as the higher energy features at both edges and support the charge-disproportionation hypothesis, and obtained an effective charge valence difference of ~0.38.

The absence of an orbital contribution to the (1/2 0 1/2) reflection is consistent with the resonant x-ray scattering signal in the $\sigma$-$\pi$' channel of the ($h0l$) type reflections at the Ni $K$ edge.[8] The signal in the $\sigma$-$\pi$' channel probes the $\langle T_1^2 \rangle$ quadrupole of the 4$p$ shell and reflects the structural deviation of the $Pbnm$ symmetry from cubic. The Ni ions have $\bar{1}$ site symmetry and the $e_g$ states are therefore non degenerate above $T_{MI}$. It has been argued that this energy splitting may be small and the degeneracy could be removed at $T_{MI}$. The asphericity (quadrupole) observed in the 4$p$ shell implies that the 3$d$ shell will have a similar asphericity though the contribution to the scattering may be insignificant. In other words, above $T_{MI}$ the orbitals are already ordered coherently with the rotation of the oxygen octahedra. Moreover, the unusual magnetic ordering wavevector is caused by the non-collinear antiferromagnetic structure, and no up-up-down-down spin orientation is present. The non-collinear magnetic structure leads to six equivalent exchange paths, which can be either ferro- or antiferromagnetic. Therefore, the magnetic structure contradicts the presence of orbital ordering with the wavevector of (1/2 0 1/2).

In conclusion, the resonant soft x-ray scattering experiments of the *2p-3d* transition of Ni probe directly the magnetic ordering of the Ni moments. The observed large branching ratio of the magnetic scattered signal is similar to that for the manganites. Calculations show, that the two observed features in the spectra can be associated with $Ni^{2+}$ and $Ni^{3+}$ contributions, and a charge disproportionation in agreement with resonant hard x-ray diffraction is obtained. The azimuthal angle dependence shows that the (1/2 0 1/2) reflection is of solely magnetic origin without any orbital contribution. Polarization analysis of the scattered beam confirms this interpretation. In contrast to earlier neutron diffraction studies, which could not unambiguously determine the magnetic structure, the resonant soft x-ray scattering results show that the magnetic structure is non-collinear. Therefore, the nickelates are materials for which orbital ordering is absent at $T_{MI}$. The metal-insulator transition is solely driven by charge disproportionation.

We thank S. W. Lovesey, J. Fernández Rodríguez, N. Binggeli and M. Altarelli for helpful discussions and remarks and the beamline staff of X11MA for its excellent support. This work was supported by the Swiss National Science Foundation and performed at SLS of the Paul Scherrer Institut, Villigen PSI, Switzerland.

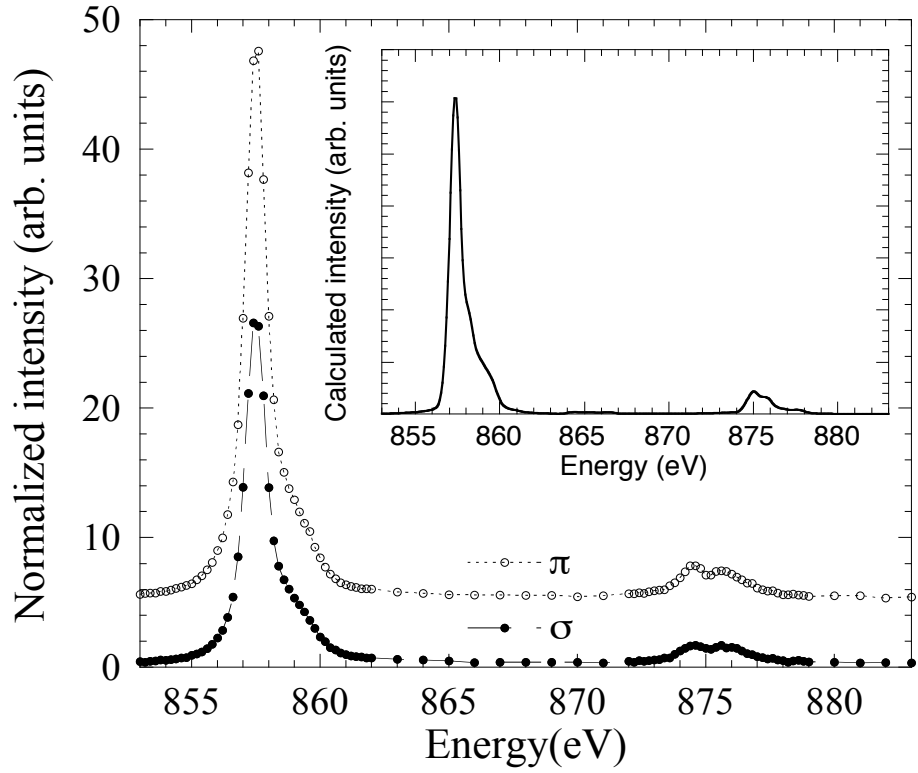

Figure 1: Absorption corrected intensity of the (1/2 0 1/2) reflection of NdNiO$_3$ at 30K for incident σ and π polarization in the vicinity of the Ni $L_{2,3}$ edges ($\psi=0°$). The incident π data are offset for clarity. The inset shows the theoretical fit using two different Ni sites.

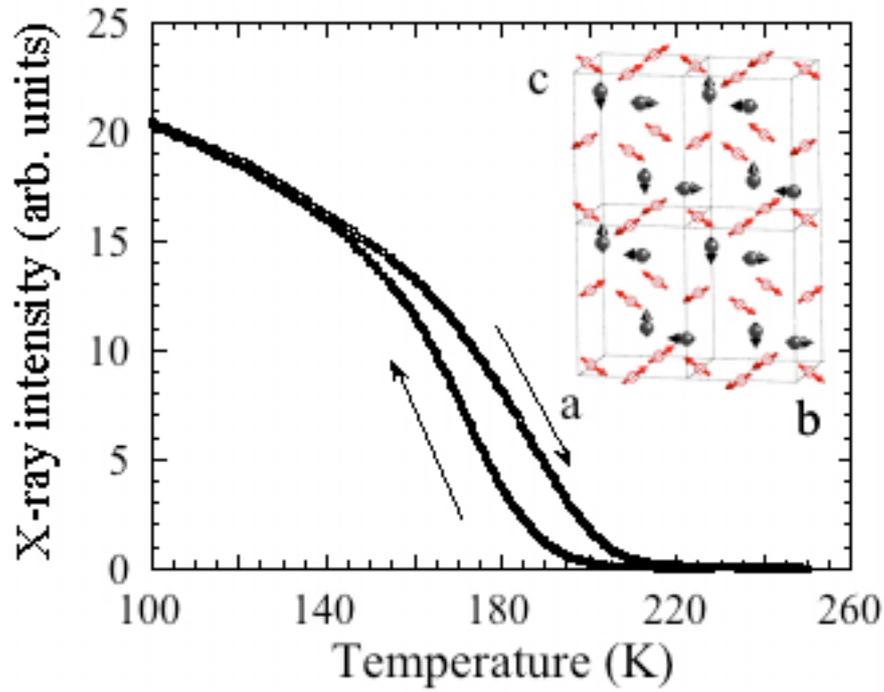

Figure 2: Temperature dependence of the (1/2 0 1/2) reflection of NdNiO$_3$ for cooling and heating taken at the Ni $L_3$-edge (857.4 eV) with π incident radiation at $\psi=90°$. Inset: Proposed magnetic structure of NdNiO$_3$. The open circles reflect the Ni ions with their magnetic dipole moments and the closed circles reflect the Nd ions with their corresponding predicted induced moments.

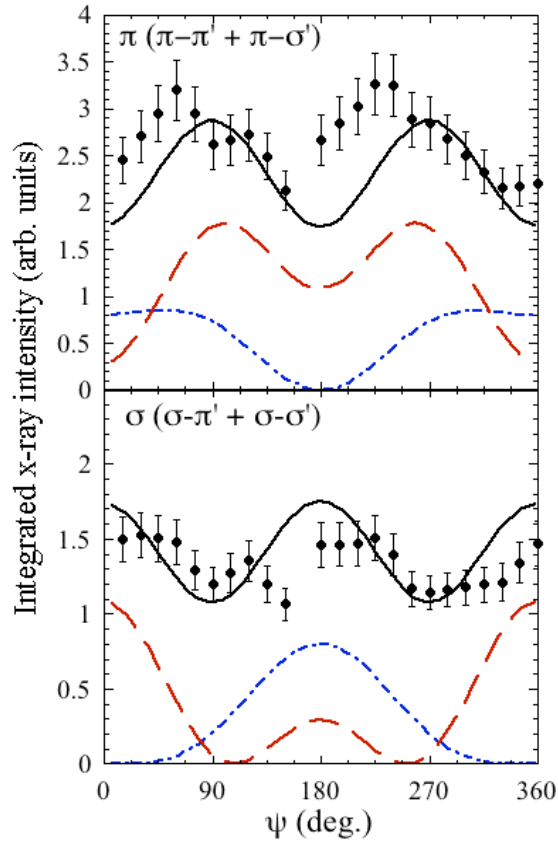

Figure 3. (colour online) Integrated intensity of the (1/2 0 1/2) reflection at T=30K as a function of azimuthal angle taken at 857.4 eV ($L_3$ edge) with π (upper panel) and σ (lower panel) incident radiation. The solid line corresponds to the calculations of the magnetic model shown in Fig. 2. The dotted and dashed lines represent the collinear models with moments along the a axis[5] and within the (a,c) plane,[6,11] respectively. $\psi=0°$ is for [010] along the z azis.

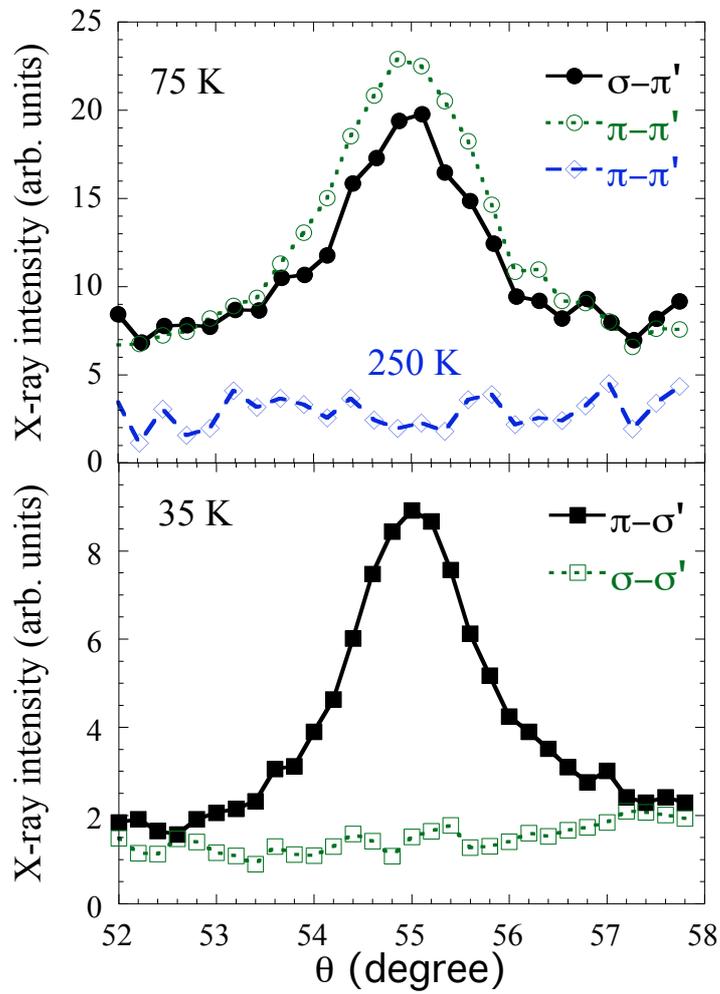

Figure 4. (colour online) Angular dependence (θ:2θ scan) of the (1/2 0 1/2) reflection at $\psi=90°$ taken for all four different polarization channels, which are taken in two different experiments. They are taken at the maximum intensity of the Ni $L_3$ edge (857.4 eV). There is a constant added to the data taken at 75K for better visibility.